# A model for exploring bird morphology

Shyamal Lakshminarayanan

261, 9[th] Cross, Tatanagar
Bangalore 560092
Lshyamal@gmail.com

**Abstract**
**A simplified model of the bird skeleton along with elongation parameters for the flight feathers is used to explore the diversity of bird shapes. Varying the small number of parameters simulates a wide range of observed bird silhouettes. The model may serve to examine developmental factors involved in these forms, help museum curators develop computational approaches to bird morphometry and has applications in computer generated illustration.**

As early as 1917, attempts were made to find simple models to explain the wide range of biological forms.[1] The modern field of morphogenesis studies the underlying biochemical and physical chain of events that lead to the formation of structures. While morphogenesis studies today principally use the tools of molecular biology, early studies were based on mathematical theory such as that proposed by Alan Turing to explain plant structures.[2] Later computer based simulation approaches have been used to study the diversity of mollusc form and colour,[3] plant structures,[4] mammalian coat patterns,[5] butterfly wing patterns[6] and the structure and colouration of bird feathers.[7][8]

In this study, a simplified model of the bird skeleton is used with a set of parameters. A program to generate a bird outline based on the given parameters was written and used to visually explore the possible variations. The taxonomic orders of birds show variations in the numbers of primaries, secondaries and tail feathers. The lengths of these feathers vary within limits across species and populations. These measurements are traditionally used in museum diagnosis for the identification of species and geographic races. The positions of the skeletal elements are highly variable and are controlled by muscles to control the flight of birds.

The simplified model of the bird skeleton along with the parameters is indicated in figure 1. The feathers are indicated as lines and their lengths are based on their position. The lengths of the feathers change with distance from body axis according to an elongation factor. The elongation factor is an exponential term and its effect on tail shapes is shown in figure 2.

Since actual morphometric data was not available, the parameter values were set by eye to reflect recognizable bird shapes. A range of silhouettes are shown in figure 3 and their corresponding parameter values are provided in table 1. The computer program (for Microsoft Windows) for generating these shapes is available for download at http://www.geocities.com/muscicapa/abob.htm and the source code is available on request.

**Acknowledgements**

I thank Drs. Richard Prum, Yale University and N. V. Joshi, Indian Institute of Science for comments on an earlier version and encouragement.

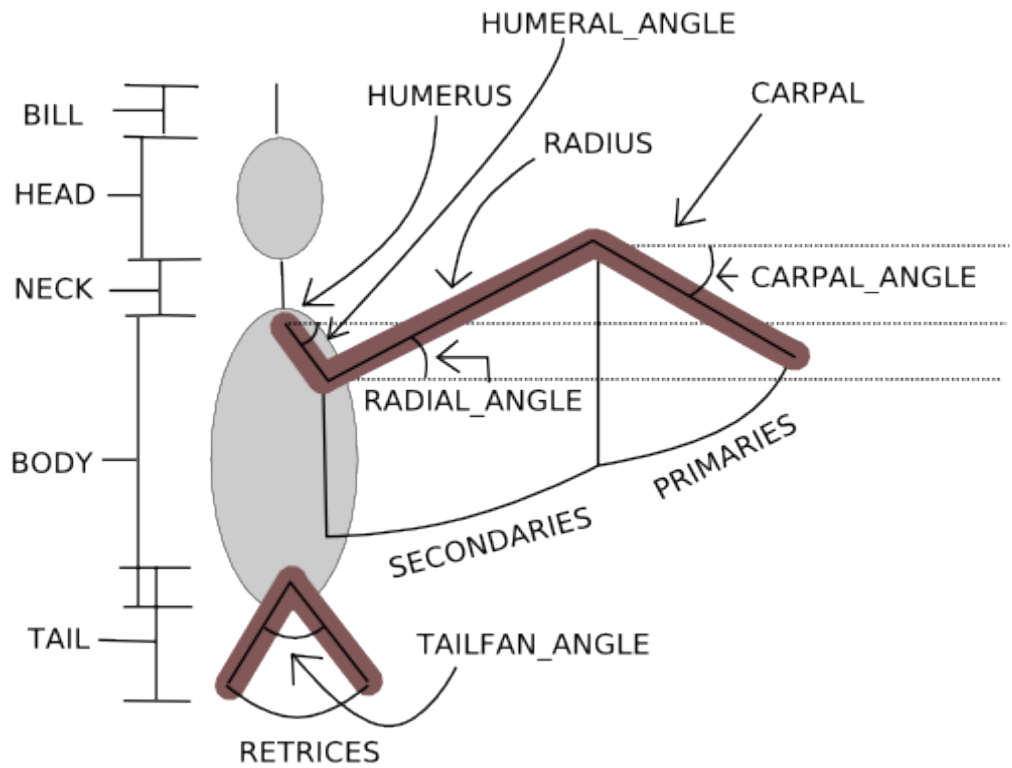

Figure 1: Parameters used to describe the structure of the model bird

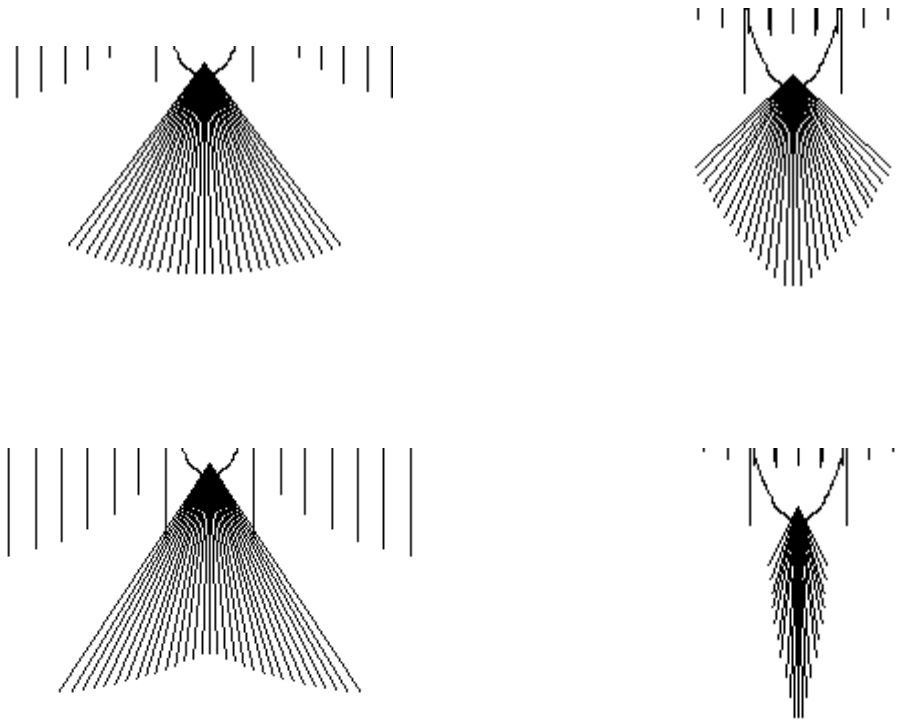

Figure 2: Elongation factors and tail shape variation
(Top-left) 1.0 – rounded tail (*Buteo*), (Top-right) 0.5 – wedge tal (*Gypaetus*)
(Lower-left) 1.5 – forked tail (*Milvus*), (Lower-right) 0.3-graduated tail (*Pica*)

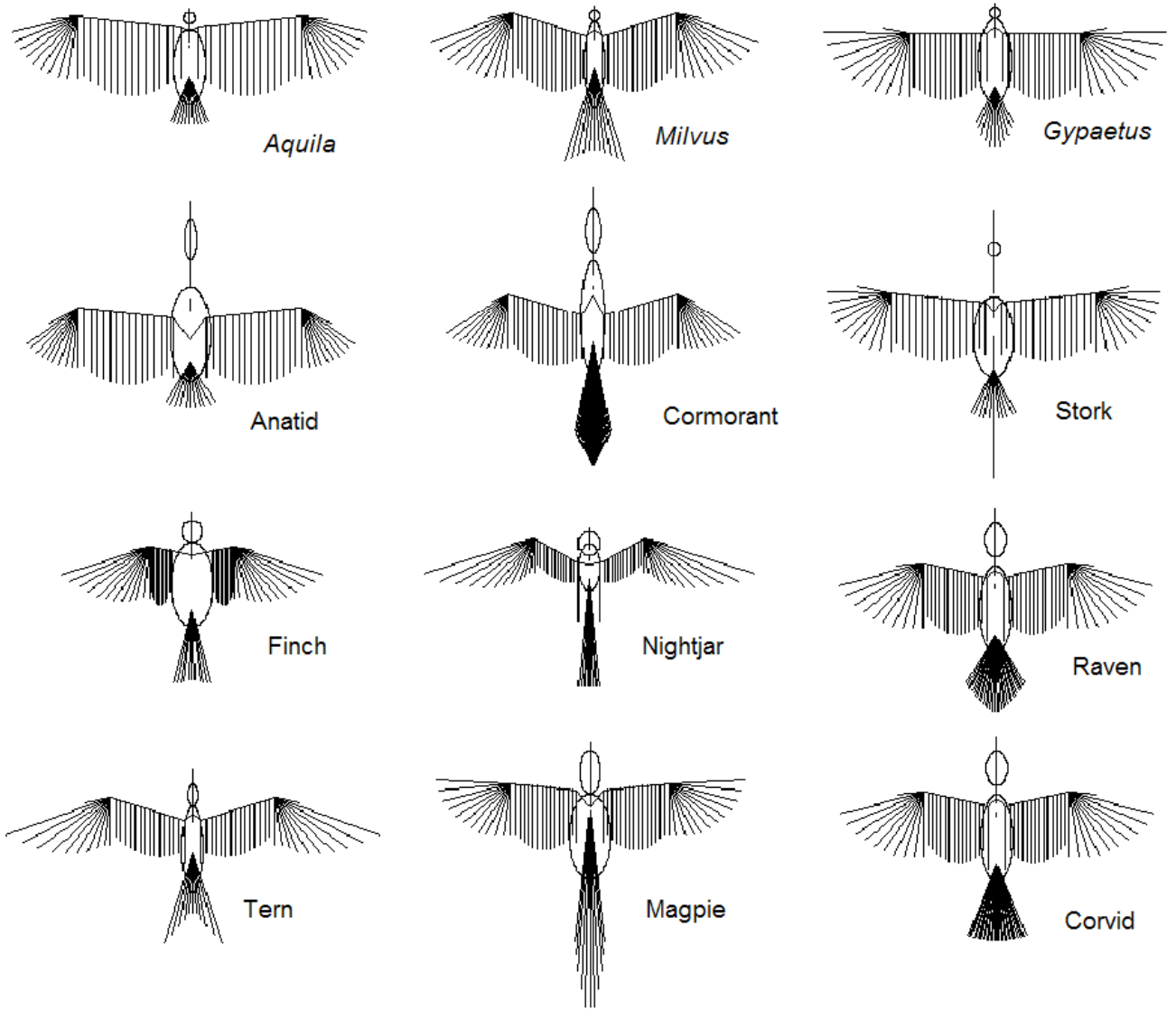

Figure 3: Silhouettes created by varying model parameters

Table 1: Parameter values used in the model to generate the silhouettes in Figure 3

| | Length | Width | Head length | Head width | Neck length | Bill length | leg length | Tail length | Tailfan angle | Tail feathers | Elongation | Tailbase | Primaries | Secondaries | Wingbase | Primary Length | Elongation | Secondary Length | Elongation | Radius length | Radius angle | Carpal length | Carpal angle | Humerus length | Humeral angle |
|---|---|---|---|---|---|---|---|---|---|---|---|---|---|---|---|---|---|---|---|---|---|---|---|---|---|
| *Aquila* eagle | 70 | 30 | 10 | 10 | 5 | 10 | 0 | 110 | 0.685 | 10 | 1 | -0.3 | 10 | 15 | 0.9 | 120 | 1.05 | 100 | 1.4 | 190 | 0.1 | 50 | -0.1 | 20 | 0.6 |
| Cormorant | 160 | 30 | 60 | 20 | 10 | 60 | 0 | 400 | 0.285 | 30 | 0.58 | -0.4 | 10 | 15 | 0.4 | 150 | 1.33 | 120 | 1.4 | 190 | 0.3 | 50 | -0.5 | 60 | -1.1 |
| Corvid | 80 | 30 | 30 | 20 | 10 | 30 | 0 | 190 | 0.585 | 30 | 1 | -0.6 | 10 | 15 | 0.9 | 120 | 1.43 | 120 | 1 | 120 | 0.2 | 50 | -0.1 | 20 | -0.6 |
| Anatid | 70 | 30 | 30 | 10 | 20 | 30 | 0 | 110 | 0.7 | 10 | 0.99 | -0.6 | 10 | 15 | -0.1 | 90 | 1.05 | 15 | 1.05 | 140 | 0.1 | 50 | -0.5 | 40 | 0.9 |
| *Milvus* | 80 | 20 | 10 | 10 | 0 | 10 | 0 | 160 | 0.6 | 10 | 1.5 | -0.1 | 10 | 15 | 0.8 | 120 | 1.43 | 90 | 1.4 | 140 | 0.3 | 50 | -0.2 | 20 | -0.6 |
| *Gypaetus* | 80 | 30 | 10 | 10 | 0 | 10 | 0 | 160 | 0.6 | 10 | 0.5 | -0.6 | 10 | 15 | 0.8 | 120 | 1.43 | 90 | 1.4 | 140 | 0 | 50 | 0.2 | 20 | -0.6 |
| Magpie | 40 | 20 | 20 | 10 | 0 | 10 | 0 | 160 | 0.3 | 10 | 0.3 | 0.6 | 12 | 15 | 0.7 | 50 | 1.5 | 50 | 1.2 | 60 | 0.1 | 0.2 | 0.2 | 20 | 0.9 |
| Nightjar | 20 | 10 | 10 | 10 | -5 | 5 | 0 | 100 | 0.2 | 10 | 1 | -0.5 | 12 | 15 | 0.1 | 30 | 4 | 50 | 0.5 | 40 | 0.5 | 20 | -0.2 | 10 | 0.2 |
| Raven | 80 | 30 | 30 | 20 | 10 | 30 | 0 | 190 | 0.785 | 30 | 0.58 | -0.6 | 10 | 15 | 0.9 | 120 | 1.43 | 120 | 1 | 120 | 0.2 | 50 | -0.1 | 20 | -0.6 |
| Finch | 40 | 20 | 10 | 10 | 0 | 10 | 0 | 90 | 0.4 | 10 | 1.1 | -0.6 | 12 | 15 | 0.7 | 50 | 1.9 | 50 | 1.1 | 20 | 0.2 | 20 | -0.2 | 20 | 0.2 |
| Tern | 80 | 20 | 20 | 10 | 0.1 | 30 | 0 | 120 | 0.6 | 10 | 2.9 | -0.1 | 10 | 15 | 0.8 | 90 | 2.8 | 80 | 1.1 | 140 | 0.3 | 50 | -0.2 | 20 | -0.6 |
| Stork | 60 | 30 | 10 | 10 | 30 | 50 | 210 | 120 | 0.6 | 10 | 1 | -0.8 | 10 | 15 | 0.6 | 90 | 1.05 | 80 | 1.2 | 140 | 0.1 | 50 | 0.2 | 20 | 0.9 |